\documentstyle[times,afd-gen,afd-orm,asymetrix,url,qb]{article}

\title{Interactive Query Formulation using 
       Query By Navigation\\
       {\normalsize Asymetrix Report 94-4}
}
\author{
   H.A. Proper\\
   Asymetrix Research Laboratory\\
   Department of Computer Science\\
   University of Queensland\\
   Australia 4072\\
   E.Proper@acm.org
}
\date{\Version}

   \input{epsf}
   \def\Figurepath{./}
   \def\Scale{0.9}
   \def\epsfsize#./##2{\Scale#./}

\begin{document}
   \maketitle
   {\sc Published as:}
\begin{quote}
  H.A.~(Erik) {Proper}. {Interactive Query Formulation using Query By Navigation}. Technical report, Asymetrix Research Laboratory, University of Queensland, Brisbane, Queensland, Australia, 1994.
\end{quote}

   \begin{abstract}
   Effective information disclosure in the context of databases with a large 
   conceptual schema is known to be a non-trivial problem. 
   In particular the formulation of ad-hoc queries is a major problem in such 
   contexts. 
   Existing approaches for tackling this problem include graphical query 
   interfaces, query by navigation, query by construction, and point to point 
   queries. 
   In this report we propose an adoption of the query by navigation mechanism
   that is especially geared towards the InfoAssistant product.
   Query by navigation is based on ideas from the information retrieval world, 
   in particular on the stratified hypermedia architecture.

   When using our approach to the formulations of queries, a user will first 
   formulate a number of simple queries corresponding to linear paths
   through the information structure. 
   The formulation of the linear paths is the result of the {\em explorative 
   phase} of the query formulation.
   Once users have specified a number of these linear paths, they may combine 
   them to form more complex queries.
   Examples of such combinations are: concatenation, union, intersection and 
   selection.
   This last process is referred to as {\em query by construction}, and is 
   the {\em constructive phase} of the query formulation process.
\end{abstract}

   \section{Introduction}

This report is concerned with an adaption of the existing idea of query by
navigation to make it fit within the InfoAssistant product.
Query by navigation has been discussed before in 
\cite{Report:92:Burgers:PSMIR}, \cite{Report:93:Proper:DisclSch}, 
\cite{PhdThesis:94:Proper:EvolvConcModels}, and \cite{Report:94:Hofstede:CSQF-QBN}.
In this report we base ourselves mainly on the latest version of the query 
by navigation mechanism as discussed in \cite{Report:94:Hofstede:CSQF-QBN}.
We have, however, stripped the existing definition from its information 
retrieval context to tailor it better to the InfoAssistant context.
 
In the previous Asymetrix reports \cite{AsyReport:94:Proper:PPQ} and
\cite{AsyReport:94:Proper:SQ}, we have already provided an elaborate
motivation for the query formulation tools we envision for InfoAssistant.
Therefore, we do not provide any further motivation for query by
navigation in this report and limit ourselves to the discussion of
the idea itself and its formal background.

The structure of this report is as follows. 
In \SRef{Sample} a sample query by navigation session is discussed, while a 
discussion of the foundations in terms of ORM and path expressions is
provided in \SRef{Basics}.
The core of this report is formed by \SRef{QBN}, in which the actual query
by navigation graph is defined.
Finally, \SRef{Concl} concludes the report.
For the reader who is unfamiliar with the notation style used in this report,
it is advisable to first read \cite{AsyReport:94:Proper:Formal}.

   \section{Exploring an Information Structure}
\SLabel{section}{Sample}

Before formally describing the notion of query by navigation, we offer the 
reader a brief example of the use of a query by navigation system.
The example shows how the system can support users when they formulate a 
query.
In our view, the process of query formulation corresponds to a search through 
the information system with the aim to gradually fulfill a user's information 
need.
Query by navigation is one of the avenues along which parts of this information
need can be formulated.
These partial formulations can then be used in the constructive phase of the 
query formulation where they are integrated into a complete query.

During query by navigation, the (partial) query of the searcher is
formulated by stepwise refining or enlarging the current description
(the {\em focus}) of this query, until the searcher recognises the current
description as the best possible description of this (part of the) query.
In the example we make use of the conceptual schema of the presidential 
database as depicted in \SRef{\PresInfStr}.
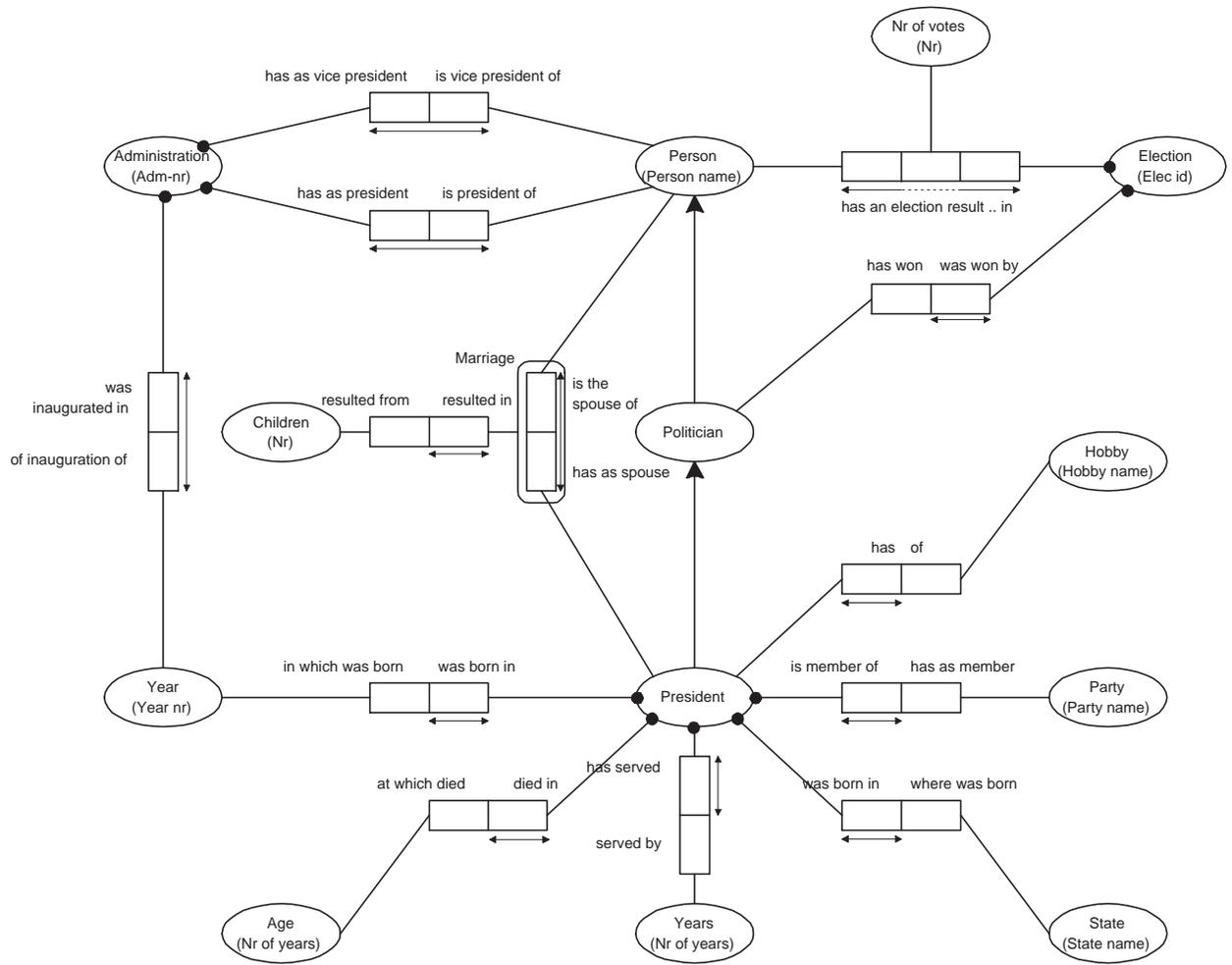
{\def\Scale{0.80} \EpsfFig[\PresInfStr]{The structure of the presidential database}}

The first node shown to the user is depicted in \SRef{\FirstExample}.
In the upper window the query by navigation dialogue is shown,
whereas the lower window displays the current query by construction session.
The query by navigation window displays the standard starting node of a 
query by navigation session; it simply lists all object types in the 
conceptual schema. 
In most Object-Role Modelling dialects and ER variations, relationship types 
can be objectified, i.e. instances of relationship types can play roles in 
other relationship types.
In the query by navigation system, relationship types are not treated as 
object types until they indeed have been objectified.
So non-objectified relationship types are not listed in the start window.

In the context of query by navigation, we use the term {\em node} 
to refer to the screens shown to the user in the query by navigation
window.
This is done to emphasize the fact that the navigation during a query by 
navigation session can be seen as the navigation through a graph (or a 
hypertext).

Each entry in a node represents one way to continue the search through the
conceptual schema.
A node thus corresponds to a moment of choice in the search process.
The order in which the alternatives are listed in the starting node, and 
nodes in general, can be based on multiple factors.
In this paper we do not discuss these factors in detail, but alternatives may 
be based on the conceptual relevance of the object types occuring in the 
alternatives (\cite{Article:94:Campbell:Abstraction}, 
\cite{AsyReport:94:Proper:PPQ}), or the user's past behaviour 
(\cite{Report:94:Berger:IRSupport}).
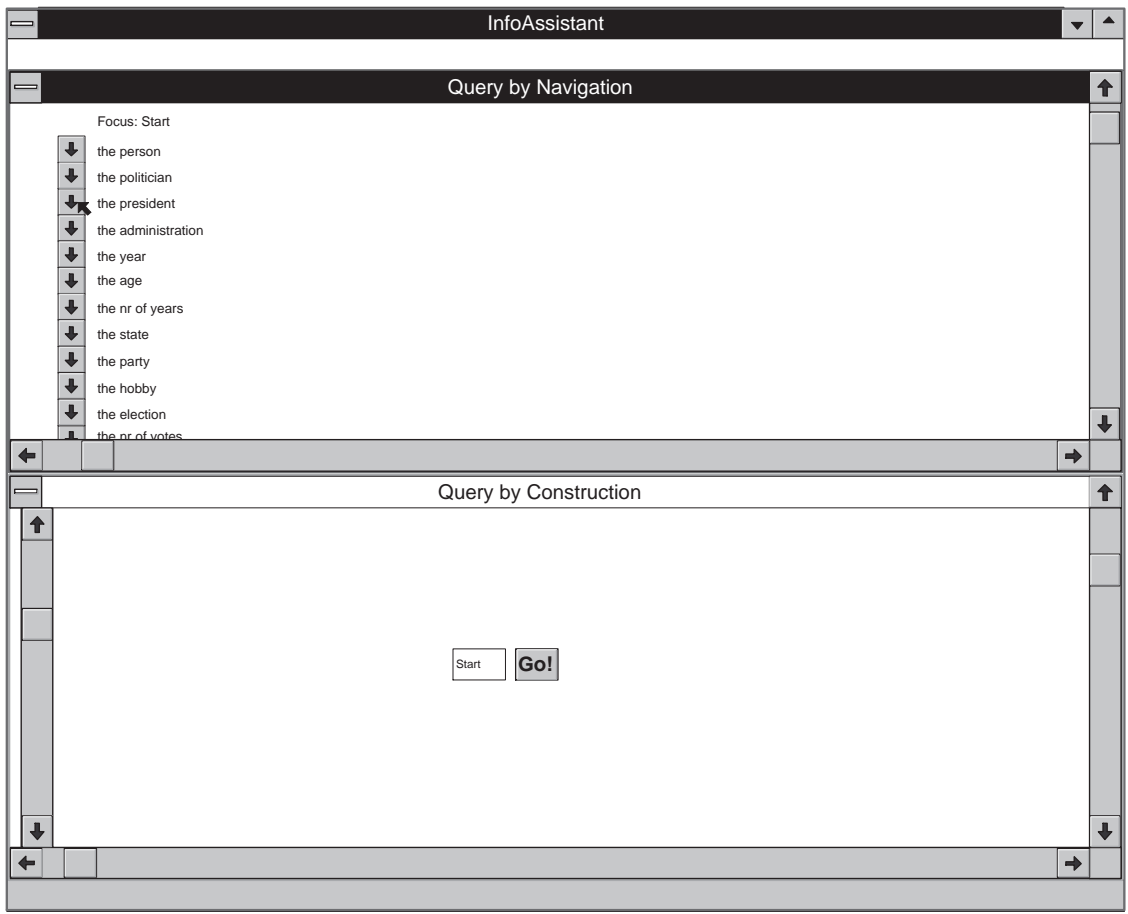
{\def\Scale{0.70} \EpsfFig[\FirstExample]{The starting node of the hyperindex}}

Let us presume the user is interested in presidents who are married and the 
number of children that resulted from these marriages.
In the starting node, the user may select `\SF{the president}' as the first 
refinement of the information need.
This leads to the example node as presented in \SRef{\SecondExample}.
The associated node shows the direct environment of type \SF{president}.
This new node contains three classes of entries. 
Firstly, the \Enlarge button takes the user back to a more general node, 
in this case the starting node.

The other class (\Refine button) represents the possible refinements of the 
current focus.
This set basically consists of the following two classes:
\begin{enumerate}
   \item For each $n$-ary relationship type in which the current focus 
         (\SF{the president}) plays a role, we have  $n-1$ possible 
         refinements since there are $n-1$ possible ways to continue the 
         path {\em through} this relationship type.

   \item Each role leading into an objectified relationship type (e.g. 
         \SF{Marriage}), or a non-binary relationship type, also results in 
         a possible refinement.
\end{enumerate}
The second class is needed to cater for the traversal of objectified 
relationship types (in the direction of the objectification), and 
furthermore, to be able to split paths on $n$-ary relationship 
types (e.g. into $n-1$ paths).

The last group of entries in the node (the \Assoc buttons) are the 
associative links.
They are derived from the subtyping hierarchy in the conceptual schema.
In our example ORM schema these are the supertypes of the 
\SF{President} object type, being \SF{Politician} and \SF{Person}.
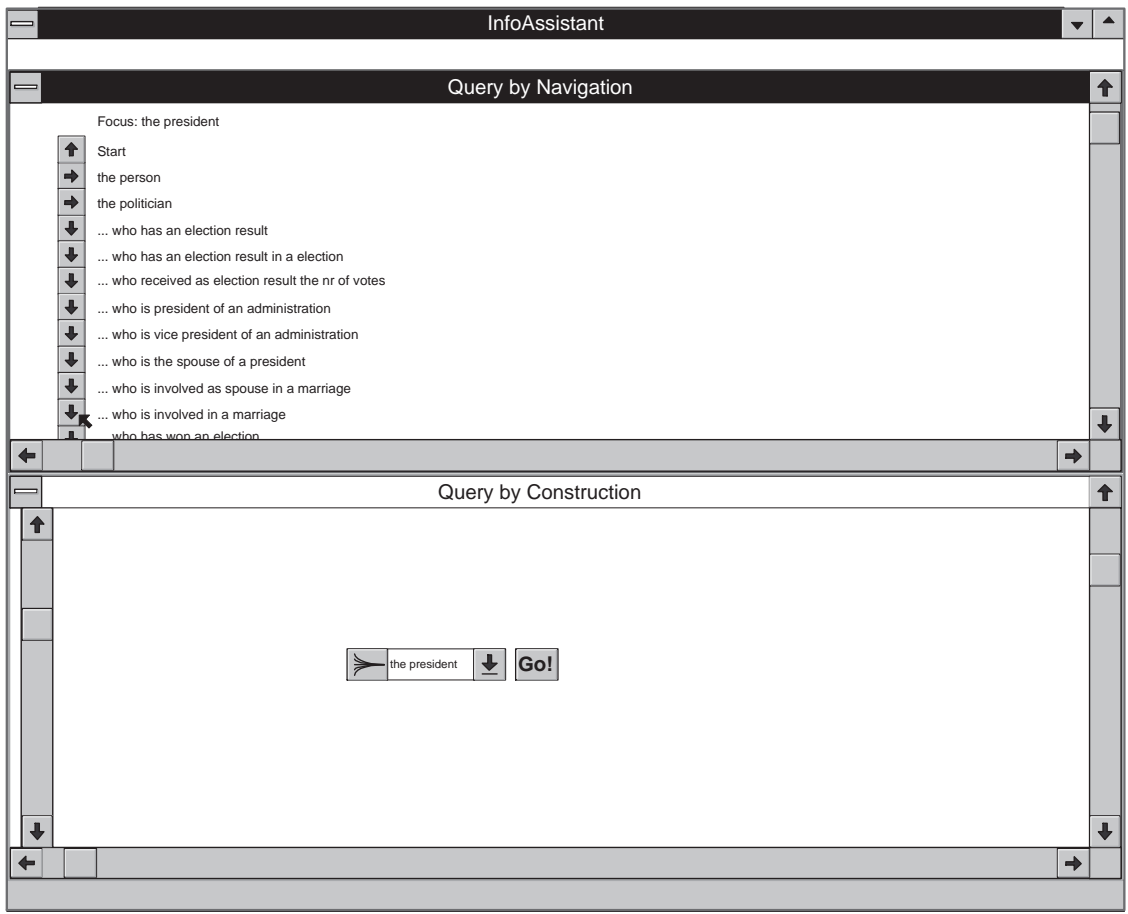
{\def\Scale{0.70} \EpsfFig[\SecondExample]{The quest for a president who is 
                                           married to a person}}

The searcher selects \SF{the president who is involved in a marriage} as the 
next focus, i.e. the objectified relationship type is the next point for 
possible further refinement.
In the resulting node the searcher continues with \SF{president involved
in marriage with person}.
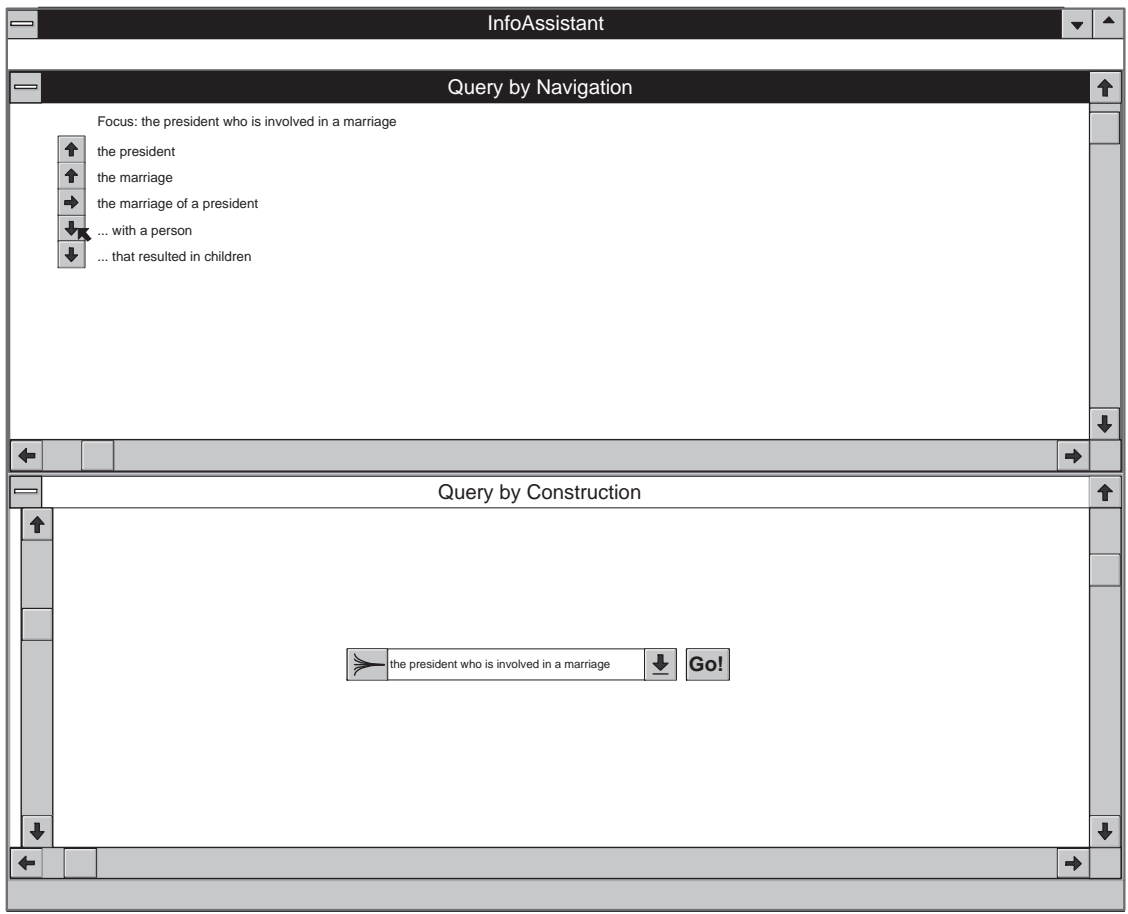
{\def\Scale{0.70} \EpsfFig[\ThirdExample]{Focus on the objectified marriage 
                                          relationship}}

The user then decides to select the refinement \SF{with a person}. 
So effectively, the user has traversed an (objectified) binary relationship 
type in two steps. 
(Note that the system verbalised this traversal in a briefer format).
The first step brought the user to the \SF{Marriage} relationship type, while 
the second step brought the user to the \SF{Person} object type. 
This leads to the node depicted in \SRef{\FourthExample}.
Note that the user could just as well have selected \SF{the president who has 
as spouse a person} in the node depicted in \SRef{\SecondExample}.

The user decides that the current focus is, for the moment, a proper 
description of the information need.
To get an impression of query the result so far the user can select the 
\SF{GO!} button. 
This should result in a window showing the result of the path formulated in
the query by navigation session.
{\def\Scale{0.70} \EpsfFig[\FourthExample]{Preleminary result in the hyperindex}}

In \cite{Report:92:Burgers:PSMIR}, \cite{Report:93:Proper:DisclSch}, 
\cite{PhdThesis:94:Proper:EvolvConcModels}, and \cite{Report:94:Hofstede:CSQF-QBN},
the query by navigation system also included the population.
In such a system, a user can either navigate through the population of the
schema, or through the conceptual schema (the type level) as discussed in 
this section.
When the query by navigation mechanism, restricted to the type level as we
propose for InfoAssistant, turns out to be successful we can indeed consider 
introducing query by navigation on a population level as well.
We realise that the navigation through the population for the formulation
of queries is only useful in the context of small database populations.
However, navigation through a hyperbase has more uses.
Firstly, it can be used as a mechanism for relevance feedback during the 
query formulation process, since it allows the user to probe the result of
the query.
Secondly, it is quite possible to use navigation through the hyperbase as a 
mechanism for the verification of conceptual schemas. 
In the design procedure for Object Role Modelling techniques example 
populations play an important role.
If a CASE tool allows for the storage of a complete example population, 
navigating through this population might turn out to be a good tool for 
the validation by the future users of the conceptual schema.

   \section{The Basics}
\SLabel{section}{Basics}

To formally define the query by navigation mechanism, we first have to 
formally define what an ORM schema is and, even more importantly, the 
(linear) path expressions need to be introduced as they form the backbone
of the query by navigation system.
We base ourselves on the formalisation of the path
expressions as provided in \cite{Report:91:Hofstede:LISA-D}, and the 
formalisation of ORM given in \cite{Report:94:Halpin:ORMPoly}.

\subsection{ORM Schemas}
A conceptual schema is presumed to consist of a set of types $\Types$. 
Within this set of types two subsets can be distinguished: the relationship 
types $\RelTypes$, and the object types $\ObjTypes$. 
Furthermore, let $\Preds$ be the set of roles in the conceptual schema. 
The fabric of the conceptual schema is then captured by two functions and 
two predicates. 
The set of roles associated to a relationship type are provided by the 
partition: 
   $\Roles: \RelTypes \Func \Powerset(\Preds)$. 
Using this partition, we can define the function $\Rel$ which returns for 
each role the relationship type in which it is involved:
   $\Rel(r) = f \iff r \in \Roles(f)$.
Every role has an object type at its base called the player of the role, 
which is provided by the function: $\Player: \Preds \Func \Types$. 
Subtyping and polymorphy of object types are captured by the predicates 
   $\Spec \subseteq \ObjTypes \Carth \ObjTypes$ and 
   $\Poly \subseteq \ObjTypes \Carth \ObjTypes$ 
respectively. 
From $\Spec$ and $\Poly$ we can derive the more general $\IdfBy$ relation,
capturing inheritance of properties between object types, by the following
derivation rules:
\begin{enumerate}
   \item $x \Spec y ~\vdash~ x \IdfBy y$
   \item $x \Poly y ~\vdash~ x \IdfBy y$
   \item $x \IdfBy y \IdfBy z ~\vdash~ x \IdfBy z$
\end{enumerate}
Using $\IdfBy$ we can define the notion of type relatedness: $x \TypeRel y$ 
for object types $x$ and $y$.
This notion captures the intuition that two object types may share instances.
This relation is defined by the following four derivation rules:
\begin{enumerate}
   \item $x \in \Types ~\vdash~ x \TypeRel x$
   \item $x \IdfBy y ~\vdash~ x \TypeRel y$  
   \item $x \TypeRel y ~\vdash~ y \TypeRel x$
   \item $x \TypeRel y \TypeRel z ~\vdash~ x \TypeRel z$
\end{enumerate}
Note that when using ORM with  advanced concepts 
(\cite{Report:91:Hofstede:LISA-D}, \cite{Report:94:Halpin:ORMPoly}) such as 
sequence types, set types, etc., the definition of $\TypeRel$ needs to be 
refined.

\subsection{Linear Path Expressions}
The central aspect of query by navigation are the (linear) path expressions
(\cite{Report:91:Hofstede:LISA-D}).
These expressions are build from (object) types, roles, and instances.
For query by navigation in InfoAssistant we only consider (object) types and 
roles.
In a linear path expressions, all these components are interpreted as 
binary relations, and can as such be concatenated to each other.

A type $t$ occurring in a path expressions corresponds to a binary 
relationship with tuples $\tuple{x,x}$ for every instance $x$ of type $t$.
A role $r$ corresponds to a binary (multiset) relationship connecting 
$\Player(r)$ to $\Rel(r)$, with tuples $\tuple{x,y}$ where $x$ is the $r$ 
part of relationship instance $y$ of $\Rel(r)$.
To traverse relationship types using a path expression, it must be possible 
to reverse the order of the $\Player(r)$ and $\Rel(r)$ part of a role. 
Therefore, $\Rev{r}$ represents the reversed binary relation associated to 
role $r$.
Note that in a forthcoming Asymetrix research report the path expressions
will be discussed in more detail.
For the moment, however, refer to \cite{Report:91:Hofstede:LISA-D} for an
elaborate formal definition, and to \cite{Report:92:Hofstede:LISA-DPromo}
for a more detailed informal discussion.

When displaying linear path expressions in the nodes of the query by 
navigation mechanism, the linear path expressions need to be verbalised. 
These verbalisations can be derived from the names given to the types and 
roles from the conceptual schema.
In this report we simply presume the existence of a function $\rho$ 
verbalising these linear path expressions.
For a more detailed discussion on the verbalisation of linear path 
expressions refer to \cite{PhdThesis:94:Proper:EvolvConcModels}.

   \section{The Query By Navigation Graph}
\SLabel{section}{QBN}

In this section we define the query by navigation graph itself.
Formally, a query by navigation graph is introduced as a structure
$\QBNGraph ~\Eq~
   \tuple{\Nodes, \Structure, \AssLinks}$.
The components of this graph are introduced below.
$\Nodes$ is the set of nodes of the graph.
Two classes of edges for the query by navigation graph are introduced. 
The first ones ($\Structure$) are the ones that are based on the structure
of the linear path expressions, whereas $\AssLinks$ provides the associative 
connections that are induced by the type relatedness of types.

\subsection{Structure based navigation}
The set of nodes of the query by navigation graph is defined by means
of a set of grammar (context-free production rules).
This grammar contains for each type $x$ a corresponding non-terminal 
(syntactic category) $\tuple{P_x}$.
Instantiations of syntactic category $\tuple{P_x}$ describe simple
properties of (instances of) type $x$,
i.e., properties that can be derived via a linear path expression starting 
in object type $x$.

The first rule of the grammar defines what a linear path expression is:
\[ \tuple{PE} ~\rightarrow~ \tuple{P\Sub{x}} \]
For any $x \in \ObjTypes$ we now have the following rules:
\[ \tuple{P\Sub{x}} ~\rightarrow~ x \]
The inheritance of properties between types leads to the following rules.
If $x \TypeRelEq y$, then:
\[ \tuple{P\Sub{x}} ~\rightarrow~ \tuple{P\Sub{y}} \]
which means that properties about $y$ may be used in expressions about $x$.
Each role $r \in \Preds$ such that $\Rel(r) \in \ObjTypes$ leads to the 
following rules:
\[ \tuple{P\Sub{\Rel(r)}} ~\rightarrow~ \tuple{P\Sub{\Player(r)}} \Conc r       \Conc \Rel(r)    \]
\[ \tuple{P\Sub{\Player(r)}} ~\rightarrow~ \tuple{P\Sub{\Rel(r)}} \Conc \Rev{r} \Conc \Player(r) \]
Finally, for all roles $r,q$ such that $r \neq q$ and $\Rel(p) = \setje{r,q}$
we have:
\[ \tuple{P\Sub{\Player(q)}} ~\rightarrow~ 
   \tuple{P\Sub{\Player(r)}} \Conc r \Conc \Rel(r) \Conc \Rev{q} \Conc \Player(q) \] 
The set of nodes in the query by navigation graph ($\Nodes$) is now formed 
by the set of linear path expressions which can be formed by the above rules (starting 
from $\tuple{PE}$), augmented with the empty path expression $\epsilon$ (which serves as 
the default starting point of a query by navigation session).
Note that the above syntax describes meta-rules,
which are concretized by substituting an actual object type
for meta-nonterminal $x$ or role $q$, $r$.
So, basically this grammar is a two level grammar
(\cite{Book:76:Wijngaarden:Algol68}).

When navigating through the query by navigation graph, one actually 
actually navigates by refinement or enlargement of a linear path expression, 
or by association.

Formally, the refinement/enlargement structure $\Structure$ of a query by 
navigation graph is a subset of $\Nodes \Carth \Nodes$.
This set is identified by the following kinds of structural links:
\begin{enumerate}
   \item a link from the empty path expression $\epsilon$ to 
         any molecule $x$. 
   \item a link from a molecule $P~x$ to a molecule $P~x \Conc r \Conc
         \Rel(r)$ if $x \TypeRelEq \Player(r) \land \Rel(r) \in \ObjTypes$.
   \item a link from a molecule $P~x$ to a molecule 
         $P~x \Conc \Rev{r} \Conc \Player(r)$ if 
         $x \TypeRelEq \Rel(r) \land \Rel(r) \in \ObjTypes$.
   \item a link from a molecule $P~x$ to a molecule 
         $P~x \Conc r \Conc \Rel(r) \Conc \Rev{q} \Conc \Player(q)$\Eol
         if $r \neq q \land \Rel(r) = \setje{r,q}$.
\end{enumerate}
where $P$ is any linear path expression, and $q,r$ are roles and $x$ is an 
object type.

\subsection{Association based links}
In our running example, each president is also a politician.
Sometimes, the user may want to refine a linear path expression ending
at \SF{politician} to a linear path expression ending at \SF{President}.
This is one of (two) reasons why we introduce the so called associative 
links.
In order to avoid chaotic structures, these links are only included
for types occurring at the end of a linear path expression.
An example of such a link is a link from 
   \SF{the president who has as spouse a person}
to
   \SF{the president who has as spouse a president}
and to
   \SF{the president who has as spouse a politician}.

The front part of a linear path expressions is manipulated only
indirectly when navigating through the graph.
To manipulate the front of the path expressions explicitly, 
path reversal is offered.
Path reversal is the second form of associative links.
For example, the link from 
\SF{the president who is involved} \SF{in a marriage}
  to 
\SF{the marriage of a president}
is a path reversal.

As stated before, we introduce the associative links ($\AssLinks$) of the 
hyperindex to cater for the relations in the identification hierarchy, as well 
as the reversal of the current focus.
Let $x,y$ be object types, then we have the following kinds of associative 
links:
\begin{enumerate}
   \item a link from a molecule of the form $P~x$ to a molecule $P~y$ if 
         $x \TypeRel y$, capturing type relatedness.
   \item a link from molecule $P$ to molecule $\SF{Rev}(P)$ if
         $P \ne \SF{Rev}(P)$, catering for the reversal of path
         expressions.
\end{enumerate}
The reversal of a path expressions by the $\SF{Rev}$ operation is 
recursively defined as:
\begin{eqnarray*}
   \SF{Rev}(P \Conc p       \Conc x) & \Eq & x \Conc \Rev{p} \Conc \SF{Rev}(P)\\
   \SF{Rev}(P \Conc \Rev{p} \Conc x) & \Eq & x \Conc p       \Conc \SF{Rev}(P)\\
   \SF{Rev}(x)                       & \Eq & x
\end{eqnarray*}
An example of such a reversal is:
\[
   \SF{Rev}(x \Conc p \Conc f \Conc \Rev{q} \Conc y)
   =
   y \Conc q \Conc f \Conc \Rev{p} \Conc x
\]

\subsection{Presentation of molecules}
Nodes are presented on the screen by verbalising the node itself (i.c. the
linear path expression) and all nodes reachable from this node, using
$\AssLinks$ and $\Structure$.
The set of reachable nodes is called the {\em direct environment} of the 
molecule.
The presentation of a node is thus made up of:
\begin{enumerate}
   \item a verbalization of the molecule itself,
         identifying the current spot (the focus) in the hyperindex.
   \item a verbalization of each immediate ancestor,
         showing how to decompose the focus into its components,
   \item a verbalization of each immediate descendant,
         which suggests how to extend the current focus.
   \item a verbalization of each associated molecule,
         showing the related alternatives.
\end{enumerate}
The presentation of a node $n$ is formally identified as:
\begin{eqnarray*}
   \SF{Present}(n) & \Eq &
   \tuple{\rho(n),\SF{Refine}(n),\SF{Enlrge}(n),\SF{Assoc}(n)}
\end{eqnarray*}
where the direct environment of $M$ is captured by:
\begin{eqnarray*}
   \SF{Refine}(n) & \Eq & \Set{\rho(r)}{\tuple{r,n} \in \Structure}\\
   \SF{Enlrge}(n) & \Eq & \Set{\rho(e)}{\tuple{n,e} \in \Structure}\\
   \SF{Assoc}(n)  & \Eq & \Set{\rho(a)}{\tuple{M,a} \in \AssLinks}
\end{eqnarray*}
All that remains to be done with respect to the presentation of the
molecules, is a proper definition of $\rho(P)$ where $P$ is a path
expression.
This can be done by a set of derivation rules, with an associated
preference (using penalty points).
As stated before, for a more detailed discussion of such a set of 
verbalisation rules please refer to 
\cite{PhdThesis:94:Proper:EvolvConcModels}.

\subsection{Navigating through the graph}
The navigation through the graph should be clear now.
If a user selects an optional refinement/enlargement/association, the 
associated linear path expression becomes the new focus and the direct 
environment of this new focus (node) is shown on the screen.

When implementing the query by navigation mechanism, it is probably wise to
calculate the direct environment of a node dynamically.
This is needed as the number of links in $\Structure$ and $\AssLinks$ can,
for obvious reasons, be extremely large.
The above definitions of $\Structure$ and $\AssLinks$, indeed allow
for such a dynamic calculation of the direct environment of a node.

   \section{Conclusions}
\SLabel{section}{Concl}

In this report we defined a limited version of the query by navigation
mechanism that is tailored for the InfoAssistant product.
In later versions of InfoAssistant, a more complete implementation of the
query by navigation mechanism may be considered.
For instance the navigation through the population and the support of 
multiple abstraction layers (\cite{Report:94:Hofstede:CSQF-QBN}).
Furthermore, the verbalisation of (linear) path expressions is an area in
which more improvements are possible.

Finally, extensive testing of the user interface of the query by navigation 
tool is required.
A number of possible refinements of the user interface exist. 
For instance, user might prefer it if refinements based on supertypes
or subtypes are explicitly marked as such.
For instance: 
\[ \SF{The president who is (as a person) vice president of an administration} \]
instead of
\[ \SF{The president who is vice president of an administration} \]
Such a feature could be added to the system as a {\em user selectable} option.

   \AddBib{asy}
   \BIBLIOGRAPHY{alpha}

\newcommand{\etalchar}[1]{$^{#1}$}
\begin{thebibliography}{WMP{\etalchar{+}}76}

\bibitem[BHW96]{Report:94:Berger:IRSupport}
F.C. {Berger}, A.H.M.~ter {Hofstede}, and Th.P. van~der {Weide}.
\newblock {Supporting Query by Navigation}.
\newblock In R.~{Leon}, editor, {\em {Information retrieval: New systems and
  current research, Proceedings of the 16th Research Colloquium of the British
  Computer Society Information Retrieval Specialists Group}}, pages 26--46,
  Drymen, United Kingdom, EU, 1996. Taylor Graham.

\bibitem[BPW93]{Report:92:Burgers:PSMIR}
C.A.J. {Burgers}, H.A.~(Erik) {Proper}, and Th.P. van~der {Weide}.
\newblock {Organising an Information System as Stratified Hypermedia}.
\newblock In H.A. {Wijshoff}, editor, {\em {Proceedings of the Computing
  Science in the Netherlands Conference}}, pages 109--120, November 1993.

\bibitem[CH94]{Article:94:Campbell:Abstraction}
L.J. {Campbell} and T.A. {Halpin}.
\newblock {Abstraction Techniques for Conceptual Schemas}.
\newblock In R.~{Sacks--Davis}, editor, {\em {Proceedings of the 5th
  Australasian Database Conference}}, volume~16, pages 374--388, Christchurch,
  New Zealand, January 1994. Global Publications Services.

\bibitem[HP95]{Report:94:Halpin:ORMPoly}
T.A. {Halpin} and H.A.~(Erik) {Proper}.
\newblock {Subtyping and Polymorphism in Object--Role Modelling}.
\newblock {\em Data {\&} Knowledge Engineering}, 15:251--281, 1995.

\bibitem[HPW93]{Report:91:Hofstede:LISA-D}
A.H.M.~ter {Hofstede}, H.A.~(Erik) {Proper}, and Th.P. van~der {Weide}.
\newblock {Formal definition of a conceptual language for the description and
  manipulation of information models}.
\newblock {\em Information Systems}, 18(7):489--523, October 1993.

\bibitem[HPW94]{Report:92:Hofstede:LISA-DPromo}
A.H.M.~ter {Hofstede}, H.A.~(Erik) {Proper}, and Th.P. van~der {Weide}.
\newblock {A Conceptual Language for the Description and Manipulation of
  Complex Information Models}.
\newblock In G.~{Gupta}, editor, {\em {Seventeenth Annual Computer Science
  Conference}}, volume~16 of {\em Australian Computer Science Communications},
  pages 157--167, Christchurch, New Zealand, January 1994. University of
  Canterbury. ISBN 047302313

\bibitem[HPW96]{Report:94:Hofstede:CSQF-QBN}
A.H.M.~ter {Hofstede}, H.A.~(Erik) {Proper}, and Th.P. van~der {Weide}.
\newblock {Query formulation as an information retrieval problem}.
\newblock {\em The Computer Journal}, 39(4):255--274, September 1996.

\bibitem[{Pro}94a]{PhdThesis:94:Proper:EvolvConcModels}
H.A.~(Erik) {Proper}.
\newblock {\em {A Theory for Conceptual Modelling of Evolving Application
  Domains}}.
\newblock PhD thesis, University of Nijmegen, Nijmegen, The Netherlands, EU,
  1994. ISBN 909006849X

\bibitem[{Pro}94b]{AsyReport:94:Proper:PPQ}
H.A.~(Erik) {Proper}.
\newblock {Interactive Query Formulation using Point to Point Queries}.
\newblock Technical report, Asymetrix Research Laboratory, University of
  Queensland, Brisbane, Queensland, Australia, 1994.

\bibitem[{Pro}94c]{AsyReport:94:Proper:SQ}
H.A.~(Erik) {Proper}.
\newblock {Interactive Query Formulation using Spider Queries}.
\newblock Technical report, Asymetrix Research Laboratory, University of
  Queensland, Brisbane, Queensland, Australia, 1994.

\bibitem[{Pro}94d]{AsyReport:94:Proper:Formal}
H.A.~(Erik) {Proper}.
\newblock {Introduction to Formal Notations}.
\newblock Technical report, Asymetrix Research Laboratory, University of
  Queensland, Brisbane, Queensland, Australia, 1994.

\bibitem[PW95]{Report:93:Proper:DisclSch}
H.A.~(Erik) {Proper} and Th.P. van~der {Weide}.
\newblock {Information Disclosure in Evolving Information Systems: Taking a
  shot at a moving target}.
\newblock {\em Data {\&} Knowledge Engineering}, 15:135--168, 1995.

\bibitem[WMP{\etalchar{+}}76]{Book:76:Wijngaarden:Algol68}
A.~van {Wijngaarden}, B.J. {Mailloux}, J.E.L. {Peck}, C.H.A. {Koster},
  M.~{Sintzoff}, C.H. {Lindsey}, L.T. {Meertens}, and R.G. {Fisker}.
\newblock {\em {Revised Report on the Algorithmic Language ALGOL 68}}.
\newblock Springer, Berlin, Germany, EU, 1976.

\end{thebibliography}
\end{document}